\documentclass[aps, a4paper,superscriptaddress, nofootinbib, showpacs, twocolumn, 10pt]{revtex4}
\usepackage{amsmath,amssymb,bm,natbib}
\usepackage{epsfig}
\usepackage{graphicx}
\usepackage{slashed}
\usepackage{ulem}

\newcommand{\beq}{\begin{equation}}
\newcommand{\eeq}{\end{equation}}
\newcommand{\bea}{\begin{eqnarray}}
\newcommand{\eea}{\end{eqnarray}}
\newcommand{\ba}{\begin{align}}
\newcommand{\ea}{\end{align}}
\newcommand{\bfig}{\begin{figure}}
\newcommand{\efig}{\end{figure}}

\newcommand{\D}{\displaystyle}

\newcommand{\gev}{\, \text{GeV}}
\newcommand{\mev}{\, \text{MeV}}

\newcommand{\tin}{t_{\text{in}}}

\newcommand{\up}{\text{up}}
\usepackage{color}

\newcommand{\omnes}{{\cal{O}}}

\begin{document}

\phantom{}
\vspace*{-17mm}

\title{Pion electromagnetic form factor at high precision with implications to $a_\mu^{\pi\pi}$ and the onset of perturbative QCD}

\author{B.Ananthanarayan}
\affiliation{Centre for High Energy Physics,
Indian Institute of Science, Bangalore 560 012, India}
\author{Irinel Caprini}
\affiliation{Horia Hulubei National Institute for Physics and Nuclear Engineering,
P.O.B. MG-6, 077125 Magurele, Romania}
\author{ Diganta Das}
\affiliation{Department of Physics and Astrophysics, University of Delhi, Delhi 110007, India}

\begin{abstract}  We extend recently developed methods used for determining the
electromagnetic charge radius and  $a_\mu^{\pi\pi}$ to obtain
 a determination of the  electromagnetic form factor of the pion, $F_\pi^V(t)$, in several significant kinematical regions, using a parametrization-free formalism based on analyticity and unitarity, with the inclusion of precise inputs from both timelike and spacelike regions. On the unitarity cut, below the first inelastic threshold, we use the precisely known phase of the form factor, known from $\pi\pi$ elastic scattering via the Fermi-Watson theorem, and above the inelastic threshold a conservative integral condition on the modulus.  We also use as input the experimental values of the modulus at several energies in the elastic region,  where the data from $e^+e^-\to \pi^+\pi^-$ and $\tau$ hadronic decays are mutually consistent, as well as the most recent measurements at spacelike momenta. The experimental uncertainties are implemented by Monte Carlo simulations.  At spacelike values $Q^2=-t>0$ near the origin,  our predictions are consistent and significantly more precise than the recent QCD lattice calculations.  The determinations at larger $Q^2$ confirm the late onset of perturbative QCD for exclusive quantities.  From the predictions of $|F_\pi^V(t)|^2$ on the timelike axis below 0.63 GeV, we obtain the hadronic vacuum polarization (HPV) contribution  to the muon anomaly, $a_\mu^{\pi\pi}|_{\leq 0.63\gev} = (132.97\pm 0.70)\times 10^{-10}$, using input from both $e^+e^-$ annihilation and $\tau$ decay, and  $a_\mu^{\pi\pi}|_{\leq 0.63 \gev} = (132.91\pm 0.76)\times 10^{-10}$ using only $e^+e^-$ input.
 Our determinations can be readily extended to obtain such contributions in any interval
 of interest lying between $2 m_\pi$ and 0.63 GeV. 
\end{abstract}

\maketitle
\section{Introduction} \label{sec:Intro}

The electromagnetic form factor of the pion,  $F_\pi^V(t)$, defined by the matrix element
\beq\label{eq:def} 
 \langle \pi^+(p')\vert J_\mu^{\rm elm} \vert
\pi^+(p)\rangle= (p+p')_\mu F_\pi^V(t),
\eeq
where $t=q^2$ and $q=p-p'$, is a fundamental observable of the strong interactions and a sensitive probe of the composite nature of the pion. An expansion near the origin to linear order in $t$, $F^V_\pi(t)= 
1+ \langle r_\pi^2 \rangle t/6$ exhibits the electromagnetic charge radius of
the pion, which has recently been determined at high precision in  Ref.~\cite{Ananthanarayan:2017efc} by a formalism based on analyticity and unitarity with phenomenological input.
The result for the electromagnetic charge radius reads $r_\pi=\sqrt{\langle r_\pi^2 \rangle}=(0.657\pm 0.003)$ fm, which
reduced the error by a half from previous determinations.  The work was achieved
by adapting the techniques introduced in  Ref.~\cite{Ananthanarayan:2016mns},
where the two-pion contribution  $a_\mu^{\pi\pi}$ to the anomalous magnetic moment of the muon
was determined in a region where experimental data have significant lack of
agreement.  
In this work, we adapt the methods introduced in these studies
to the determination of the form factor itself in several kinematic regions of interest.
In contrast to  the prior investigations, where a single number was determined
in each of them, in the present work we obtain the values of the form factor at a large
number of points.

We recall that there is a large amount of information, both theoretical and experimental, on the pion vector form factor, making it one of the most investigated quantity in hadron physics. 
The form factor determination at high precision  is of utmost importance to several observables including the low-energy dipion contribution to the muon $g-2$, and poses a significant challenge to experiment as well as to theory. 
Theoretical studies are based  at low energies on nonperturbative approaches and effective theories 
of the type first formulated by Weinberg~\cite{Wein}, and at large energies on perturbative QCD. In the framework of chiral perturbation theory (ChPT), the effective realization of QCD at low energies first formulated at one loop order
with two \cite{Gasser:1983yg}
and three light quark flavours \cite{GaLe1,Gasser:1984ux},  the pion vector form factor has been calculated  up to two loops \cite{GaMe, CFU, BiCo, BiTa, Bijnens:1998fm}. 
Lattice gauge theory has recently become another useful nonperturbative tool for the 
calculation of the form factor at low energies \cite{Aoki:2016frl, ETM}.  

The form factor is also a probe of energies at which asymptotic QCD predictions are expected to set in.
Perturbative  QCD  predicts the behavior at large momenta along the spacelike axis, 
where  $Q^2\equiv -t\gg 0$  \cite{Lepage:1979zb}-\cite{pQCD4}. The leading order (LO) asymptotic behavior is
\beq\label{eq:qcd}
F^{V,LO}_\pi(-Q^2)\sim \frac{8 \pi F_\pi^2 \alpha_s(Q^2)}{Q^2}, \quad\quad Q^2\to\infty,
\eeq
where $F_\pi= 131 \mev$ is the pion decay constant and $\alpha_s(Q^2)=4\pi/[9\ln(Q^2/\Lambda^2)]$ is the  running strong 
coupling to one loop with three active light quark flavors. NLO corrections to (\ref{eq:qcd}) have been calculated in \cite{Melic:1998qr,Melic:1999mx}. 

The experimental information available on the pion form factor is very rich.   This quantity was measured at spacelike
values $Q^2>0$  with increasing precision from electron-pion scattering \cite{Amendolia:1986wj} and pion electro-production from nucleons \cite{Brown:1973wr,  Bebek:1974iz, Bebek:1976ww, Ackermann:1977rp, Bebek:1978, Brauel:1979zk, Volmer:2000ek, Tadevosyan, Horn, Huber}, the most precise being the recent results of 
the JLab collaboration \cite{Horn, Huber, Huber1}. On the timelike axis, for $t\ge 4 m_\pi^2$,  where the form factor is complex,  its modulus has been measured from the cross section of the process $e^+e^-\to\pi^+\pi^-$ \cite{Akhmetshin:2006bx}-\cite{Ablikim:2015orh} and, using isospin symmetry, from the  $\tau\to \pi\pi\nu\tau$ decay 
\cite{Anderson:1999ui}-\cite{Fujikawa:2008ma}.

Due to the 
extensive experimental and theoretical information, the pion vector form factor is, compared with other hadronic quantities, a well-known function. However,  the  precision does not reach the same level for all timelike and spacelike momenta.  A better precision is required on the spacelike axis, for checking the consistency with experimental data and for testing the calculations provided by lattice QCD at low momenta and perturbative QCD at larger momenta.  On the timelike axis, at low energies  the  phase of the form factor is well known, being equal by Fermi-Watson theorem  \cite{Fermi:2008zz, Watson:1954uc} to the $\pi\pi$ scattering  $P$-wave phase shift, which has been calculated with high precision using ChPT and Roy equations \cite{ACGL, Caprini:2011ky, GarciaMartin:2011cn}. However, the modulus is poorly known, due to the difficulties of the experimental measurements in this region: only two experiments, {\it BABAR} \cite{Aubert:2009ad} and KLOE \cite{Ambrosino:2008aa, Ambrosino:2010bv, Babusci:2012rp} 
reported data at low energies, and unfortunately they are not consistent with each other. 

This situation drastically affects the calculation of the hadronic vacuum polarization (HVP) contribution to the muon anomaly, $a_\mu=(g-2)_\mu/2$, a quantity which plays an important role for testing the standard model and finding possible signals of new physics.   The great interest in the muon anomaly is motivated by the present discrepancy  of about 3 to 4$\sigma$ between theory and experiment.  
New generation measurements of muon $g-2$  planned at Fermilab\footnote{The E989 experiment at Fermilab has started its pilot runs and is gathering data at an accelerated pace.} \cite{Venanzoni:2012qa} and JPARC \cite{JPARC} are expected to produce results with experimental errors at the level of $16 \times 10^{-11}$, a factor of 4 smaller compared to the  Brookhaven measurement \cite{Bennett:2006fi}.   This requires a precision at the same level also for the theoretical result: see for instance Ref. \cite{JPARC} for an updated review, Refs. \cite{Davier:2017, Teubner:2018} for most recent phenomenological determinations, and Ref. \cite{Blum:2018mom} for a recent lattice calculation. 

Dispersion theory, which exploits analyticity and unitarity, is a powerful tool  for performing the analytic continuation of the form factor to energies where it is not precisely known.  The pion vector form factor is an analytic function in the complex $t$ plane  cut along  the real axis for $t\ge t_+$,  where $t_+= 4 m_\pi^2$ is the first unitarity threshold. Moreover, it is normalized as  $F_\pi^V(0)=1$, and satisfies the Schwarz reflection property $F_\pi^V(t^*)=(F_\pi^V(t))^*$.  It turns out that the standard dispersion relation, based on the Cauchy integral, is not suitable for $F_\pi^V(t)$, since it requires the knowledge of its imaginary part on the unitarity cut, which is not available in a straightforward way.  On the other hand, as mentioned above, in the limit of exact isospin symmetry, the Fermi-Watson theorem \cite{Fermi:2008zz, Watson:1954uc} states that  below the first inelastic threshold,  the phase of $F_\pi^V(t)$ is equal   to the $P$-wave phase shift of  $\pi\pi$ elastic scattering, which is better known. Many dispersion analyses of the pion vector form factor have been based on the so-called Omn\`es representation,  which amounts to reconstruct an analytic function from its phase on the cut. However, this approach involves some assumptions on the phase  above the inelastic threshold, where it is not known, and on  the positions  of the possible zeros in the complex plane. A related approach uses specific parametrizations which implement the analytic properties of the form factor. Recent  analyses based on this approach are \cite{Hanhart:2017, Colangelo:2018}. 

In the present paper, we  use a method based on analyticity and unitarity for calculating the form factor in kinematical regions where it is not precisely known, using  the more precise input available in other energy regions. We implement the phase of the form factor along a part of the unitarity cut, where it is well known, and information on the modulus on the remaining part of the cut. Thus, our method is neither a standard dispersive representation, nor a  specific parametrization for the analytic extrapolation in the complex momentum plane. The advantage is that we can implement only known input, avoiding to a large extent model-dependent assumptions about the behavior of the form factor in regions where it is less known.  The price to be paid was the fact that we do not obtain definite values for the extrapolated quantity, but  only optimal allowed ranges for it, in terms of the phenomenological input. This shortcoming has been overcome now as described below.

This method, proposed in \cite{Caprini:1999ws} and presented in detail in the review \cite{Abbas:2010jc},
 has been applied already in several papers \cite{Ananthanarayan:2013dpa,Ananthanarayan:2012tn, Ananthanarayan:2012tt, Ananthanarayan:2013zua}, where optimal bounds on the pion vector form factor in various energy regions have been derived. An important improvement has been achieved  by implementing the statistical distribution of the experimental input by Monte Carlo simulations, which converted the analytic bounds into allowed intervals with definite confidence levels.
 This elaborate formalism was applied in Refs. \cite{Ananthanarayan:2016mns} and \cite{Ananthanarayan:2017efc} for the calculation with a remarkable accuracy of the low-energy HVP contribution  to muon $g-2$ and the pion charge radius, respectively. In the present paper, we  now complete the task of determining the form factor itself to equally remarkable accuracy.

The outline of this paper is as follows:  
in Sec. \ref{sec:aim}  we  review the  conditions used as input and formulate the objective of the paper as an extremal problem on a class of analytic functions.  In Sec. \ref{sec:input} we give a detailed description of the 
 information used as input, and in Sec. \ref{sec:ff} we  describe the calculation of the bounds and the Monte Carlo simulation  used for implementing the  uncertainties of the input data. In this section we also explain the prescription of combining the predictions from different experiments applied in our work.  Section \ref{sec:results} contains our results and Sec. \ref{sec:conclusion}  a summary and our conclusions.  In the Appendix, we present the solution of the functional extremal problem formulated  in Sec.  \ref{sec:aim}, which is the mathematical basis of our approach. 

\vspace{0.3cm}
\section{Extremal problem\label{sec:aim}}
Our aim is to make precision predictions for the pion vector form factor in several regions on both spacelike and timelike axis. In particular, we will be interested in the modulus $|F_\pi^V(t)|$  in the low energy region $t_+ \le t \le (0.63 \gev)^2$ where $t_+=4m_\pi^2$, which will allow a new determination of the pion-pion contribution to the muon anomaly $a_\mu$ from this region. We will determine also the form factor $F_\pi^V(t)$ in the unphysical region $0\le t\le t_+$ and at spacelike values $t<0$. 

We summarize below the conditions adopted as input. 
 We implement first the normalization imposed by gauge invariance at $t=0$, expressed  by:
\beq\label{eq:taylor}
	F_\pi^V(0) = 1.  
\eeq
An important ingredient is Fermi-Watson theorem \cite{Fermi:2008zz, Watson:1954uc} mentioned above. Since this theorem is valid in the exact isospin limit, we must first  remove the main  isospin-violating effect in the pion vector form factor, known to arise from $\omega-\rho$ interference. We shall follow standard approach \cite{Leutwyler:2002hm, Hanhart:2012wi} to do this, by defining a purely $I=1$ function $F(t)$ as
\beq\label{eq:F}
F(t)=F_\pi^V(t)/F_\omega(t),
\eeq
where $F_\omega(t)$ includes the $I=0$ contribution due to $\omega$. Then Fermi-Watson theorem writes as
\beq\label{eq:watson}
{\rm Arg} [F(t+i\epsilon)]=\delta_1^1(t),  \quad\quad t_+ \le t \le \tin,
\eeq
where $\delta_1^1(t)$ is the phase-shift of the $P$-wave of $\pi\pi$ elastic scattering and 
$\tin$ is the first inelastic threshold.

Above the inelastic threshold $\tin$, where the phase is not known, we shall use the phenomenological  information available on the modulus at intermediate energies, and perturbative QCD at high energies. Since the precision is not enough to impose the condition  at each $t$ above $\tin$, we shall adopt a weaker condition, written as
\beq\label{eq:L2}
 \D\frac{1}{\pi} \int_{\tin}^{\infty} dt \rho(t) |F_\pi^V(t)|^2 \leq  I,
 \eeq
where $\rho(t)>0$ is a suitable positive-definite weight, for which the integral  converges and an accurate evaluation of $I$ from the available information is possible. 

We shall use, in addition, the experimental value of the form factor at one  spacelike energy: 
\beq\label{eq:val}
F_\pi^V(t_s)= F_s \pm \epsilon_s, \qquad t_s<0,
\eeq
and the
 modulus at one energy in the elastic region of the timelike axis, where it is known with precision from experiment: 
\beq\label{eq:mod}
|F_\pi^V(t_t)|= F_t \pm \epsilon_t, \qquad t_+< t_t <\tin.
\eeq

The aim of our work can be expressed as the following functional extremal problem: using as input the conditions (\ref{eq:taylor})-(\ref{eq:mod}), derive optimal upper and lower bounds on $|F_\pi^V(t)|$ on the unitarity cut below $0.63 \gev$,  and on  $F_\pi^V(t)$ on the real axis  for $t< t_+$. 

The solution of the extremal problem and the algorithm for obtaining the bounds are presented for completeness in the Appendix. It will be applied in Sec. \ref{sec:ff} for making precise predictions on the form factor in the regions of interest. In Sec.~\ref{sec:input} we shall describe the phenomenological information used as input.

\section{ Input in the extremal problem\label{sec:input}}
For the function $F_\omega(t)$, which accounts for the isospin violation due to $\omega$ resonance, we shall use the parametrization\footnote{An alternative parametrization written as a dispersion relation in terms of the imaginary part of (\ref{eq:rhoomega}) leads practically to the same results.} proposed in \cite{Leutwyler:2002hm, Hanhart:2012wi}:
\begin{align}\label{eq:rhoomega}
 F_\omega(t)= 1+\epsilon\frac{t}{(m_\omega-i\Gamma_\omega/2)^2 -t},
\end{align}
with  $\epsilon=1.9\times 10^{-3}$. This function is normalized as $F_\omega(0)=1$ and, due to the small value $\Gamma_\omega=8.49 \mev$  \cite{PDG}, is highly peaked around $\sqrt{t}=m_\omega=782.65 \mev$. In our treatment, we first converted the experimental values of  $F_\pi^V(t)$ used as input  in Eqs. (\ref{eq:L2}),  (\ref{eq:val}) and  (\ref{eq:mod}) to the isospin-conserving function $F(t)$ defined in (\ref{eq:F}), solved the extremal problem for this function and finally reinserted the factor $F_\omega(t)$ in the results. Actually, since we do not include the resonance region in our study, the corrections due to $F_\omega(t)$ are very small in all the kinematical regions considered,  and are practically negligible for $t\leq 0$.

The first significant inelastic threshold $\tin$ for the pion form factor is due to the  opening of the $\omega\pi$ channel, {\it i.e.}, $\sqrt{\tin}=m_\omega+m_\pi=0.917\,\gev$. 
Below this threshold, we use in  (\ref{eq:watson}) the  
phase shift $\delta_1^1(t)$ obtained from dispersion relations and Roy equations applied to $\pi\pi$ scattering in Refs. \cite{ACGL, Caprini:2011ky} and \cite{GarciaMartin:2011cn}, 
which we denote as Bern and Madrid  phase, respectively. Actually, in the calculation of the Bern phase, for the $P$-wave phase shift some input from previous data on the form factor was used at the matching point 0.8 GeV, which may raise doubts  of a circular calculation (this problem was discussed recently also in \cite{Colangelo:2018}).  However, we note that the Bern value at 0.8 GeV is practically identical to what has been called ``constrained'' fit to data (CFD) solution of the Madrid phase  \cite{GarciaMartin:2011cn}, which we adopt, and which is independent of form factor data. Actually, the  error attached to this input to Bern phase is larger (more than double)  than the uncertainty attached to the CFD solution, which reduces the possible bias.  Moreover, as we shall explain later, in our determination we take the simple average of the results obtained with the two phase-shifts, which reduces further the potential bias produced by this input and practically avoids the danger of circularity. 

We have calculated the integral (\ref{eq:L2}) using the {\it BABAR} data \cite{Aubert:2009ad} 
from $\tin$ up to $\sqrt{t}=3\, \gev$, smoothly continued with  a constant value for the modulus 
in the range $3\, \gev \leq \sqrt{t} \leq 20 \gev$,  and  a $1/t$ decreasing modulus at higher energies, as predicted by perturbative QCD \cite{Farrar:1979aw,
Lepage:1979zb, Melic:1998qr, Melic:1999mx}.
This choice is expected to overestimate the true value of the integral (see Refs. \cite{Ananthanarayan:2013zua, Ananthanarayan:2012tn, Ananthanarayan:2012tt} for a detailed discussion), which has the effect of leading to weaker bounds  due to a 
monotonicity property discussed in the Appendix. 
As concerns the weight $\rho(t)$, several choices have been investigated in  \cite{Ananthanarayan:2013zua}, leading to stable results in most of the investigated regions. In the present work,  we have adopted the weight $\rho(t)=1/t$, for which 
the contribution of the range above 3 GeV to the integral (\ref{eq:L2}) is 
only of $1\%$. The value of $I$ obtained with this weight is \cite{Ananthanarayan:2013zua}
\beq\label{eq:Ivalue1} 
I=0.578 \pm 0.022, 
\eeq 
where the uncertainty is  due to the {\it BABAR} experimental errors. In the calculations 
we have used as input for $I$  the central value quoted in Eq. (\ref{eq:Ivalue1}) 
increased by the  error, which leads to the most conservative bounds due to the 
monotonicity property mentioned above.

On the spacelike axis at moderate and large $Q^2$ the form factor  is extracted indirectly,  from experimental measurements of the pion electro-production from a nucleon target, where a virtual photon couples to a pion in the cloud
surrounding the nucleon. As a consequence, there are uncertainties
associated with the off-shellness of the struck pion and the
consequent extrapolation to the physical pion mass pole, which
leads to uncertainties in the extraction of the
form factor. The errors appear to be under control in the most recent
determinations of  $F_\pi$ Collaboration at JLab \cite{Horn, Huber}, as shown in the subsequent analysis \cite{Huber1}. 
Therefore, as spacelike  input (\ref{eq:val}) we have used the values \cite{Horn, Huber}  
\bea\label{eq:Huber}	
F_\pi^V(-1.60\,\gev^2)= 0.243 \pm  0.012_{-0.008}^{+0.019}, \nonumber \\ 
F_\pi^V(-2.45\, \gev^2)=  0.167 \pm 0.010_{-0.007}^{+0.013}.
\eea
 
We mention that we do not use as input the data on the spacelike axis near the origin, obtained from $e\pi$ scattering by NA7 Collaboration \cite{Amendolia:1986wj}.  
We shall however compare our predictions for this region with the NA7 data and with the lattice calculations \cite{ETM}.

A major role in increasing the strength of the bounds is played by condition (\ref{eq:mod}). We shall take  $0.65\gev \leq\sqrt{t_t} 
\leq 0.71 \gev$, since in this region the modulus measured by various experiments exhibits smaller variations than in other 
energy regions and a higher degree of mutual consistency. Moreover, this region is close to  the region of interest and therefore has a stronger effect on improving the bounds than the input from higher energies. 
The $e^+e^-$ data are taken below 0.705 GeV and the $\tau$-decay data below 0.710 GeV, with the exception of
one datum from CLEO that corresponds to an energy of 0.712 GeV.  Since this last datum is at an energy
that is only marginally higher than the upper limit
of the aforementioned energy range, it is included in the analysis.

 The numbers of experimental points from various experiments, used as input in our analysis, are summarized  in Table \ref{tab:1}.  We emphasize that in this region the $e^+e^-$-annihilation and $\tau$-decay experiments are fully consistent, so it is reasonable to use  all the experiments on an equal footing.
\begin{table}[t]
\centering
\renewcommand{\arraystretch}{1.3}
\begin{tabular}{l r }
\hline
Experiment ~~~& Number of points\\
\hline
CMD2 \cite{Akhmetshin:2006bx} & 2 \\
 SND \cite{Achasov:2006vp}& 2  \\
{\it BABAR}  \cite{Aubert:2009ad,Lees:2012cj}  & 26 \\
 KLOE 2011 \cite{Ambrosino:2010bv}& 8 \\
 KLOE 2013 \cite{Babusci:2012rp}& 8 \\ 
 BESIII  \cite{Ablikim:2015orh}& 10 \\ 
\hline
CLEO \cite{Anderson:1999ui} & 3\\
ALEPH  \cite{Schael:2005am, Davier:2013sfa} & 3\\
OPAL \cite{Ackerstaff:1998yj}&3\\
Belle  \cite{Fujikawa:2008ma} &2\\
\hline
\end{tabular} 
\caption{Number of points in the region $0.65\gev \leq\sqrt{t} 
\leq 0.71\gev$ where the modulus is measured by the $e^+e^-$ annihilation and $\tau$-decay experiments considered in the analysis. \label{tab:1}}
\end{table}

The extraction of the values of timelike modulus $|F_\pi^V(t)|$ from the  cross-section of the process $e^+e^-\to\pi^+\pi^-$  and the spectral function measured in $\tau$-decay experiments requires the application of several corrections,  described in detail in Appendix B of \cite{Ananthanarayan:2016mns}. In particular, for the $e^+e^-$ experiments the isospin correction due to $\omega$ has been applied as discussed above, and the vacuum polarization has been removed  from the data.
\vspace{-0.3cm}
\section{Determination of the form factor and its uncertainty \label{sec:ff}}
Using the algorithm presented in the Appendix, we obtain an allowed range for the value of $F_\pi^V(t)$ (or  $|F_\pi^V(t)|$) at an arbitrary point $t<\tin$  for every set of specific values  of the input quantities.   However, with the exception of the exact condition $F_\pi^V(0)=1$,  the input quantities are known only with some uncertainties.
 One of the key aspects
of our calculation is the proper statistical treatment of the errors. This is achieved by randomly sampling each of the
input quantities with specific distributions: the phase of $F_\pi^V(t)$, which is the result of a theoretical calculation, is assumed to be
uniformly distributed, while for the spacelike and the timelike data, which are known from experimental measurements, we adopt Gaussian  distribution with the measured central value as mean and the quoted error (the biggest error for spacelike data where the errors are asymmetric) as standard deviation.

For each point from the input statistical sample, if the input values are compatible, we calculate from Eq. (\ref{eq:ABC}) upper and lower bounds on $F_\pi^V(t)$ (or  $|F_\pi^V(t)|$). Since all the values between the extreme points are equally allowed, we uniformly generate values of $F_\pi^V(t)$ (or  $|F_\pi^V(t)|$)  in between the bounds.  For convenience, the minimal separation between the generated points was set at $10^{-3}$ and  for  allowed 
intervals  smaller than this limit no intermediate points were created. In this way, for each input from one spacelike $t_s<0$  and one timelike  $t_t$ in the region $(0.65-0.71) \gev$, we obtain a large  sample of values of $F_\pi^V(t)$ (or  $|F_\pi^V(t)|$). The results proved to be stable against the variation of the size of the random sample and the minimal separation  mentioned above.

\begin{figure*}[thb]\label{fig:hist}\vspace{0.5cm}
	\begin{center}
		\includegraphics[width = 8.2cm]{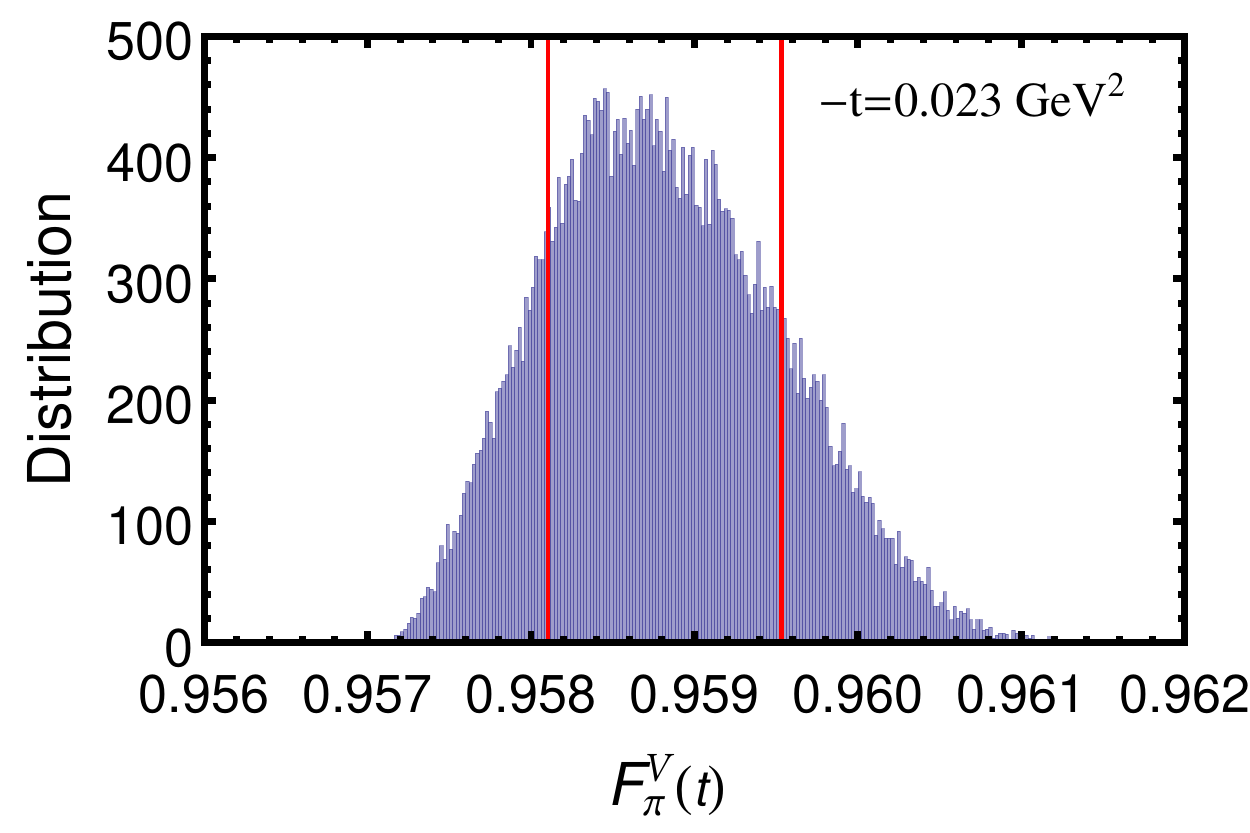}
		\includegraphics[width = 8cm]{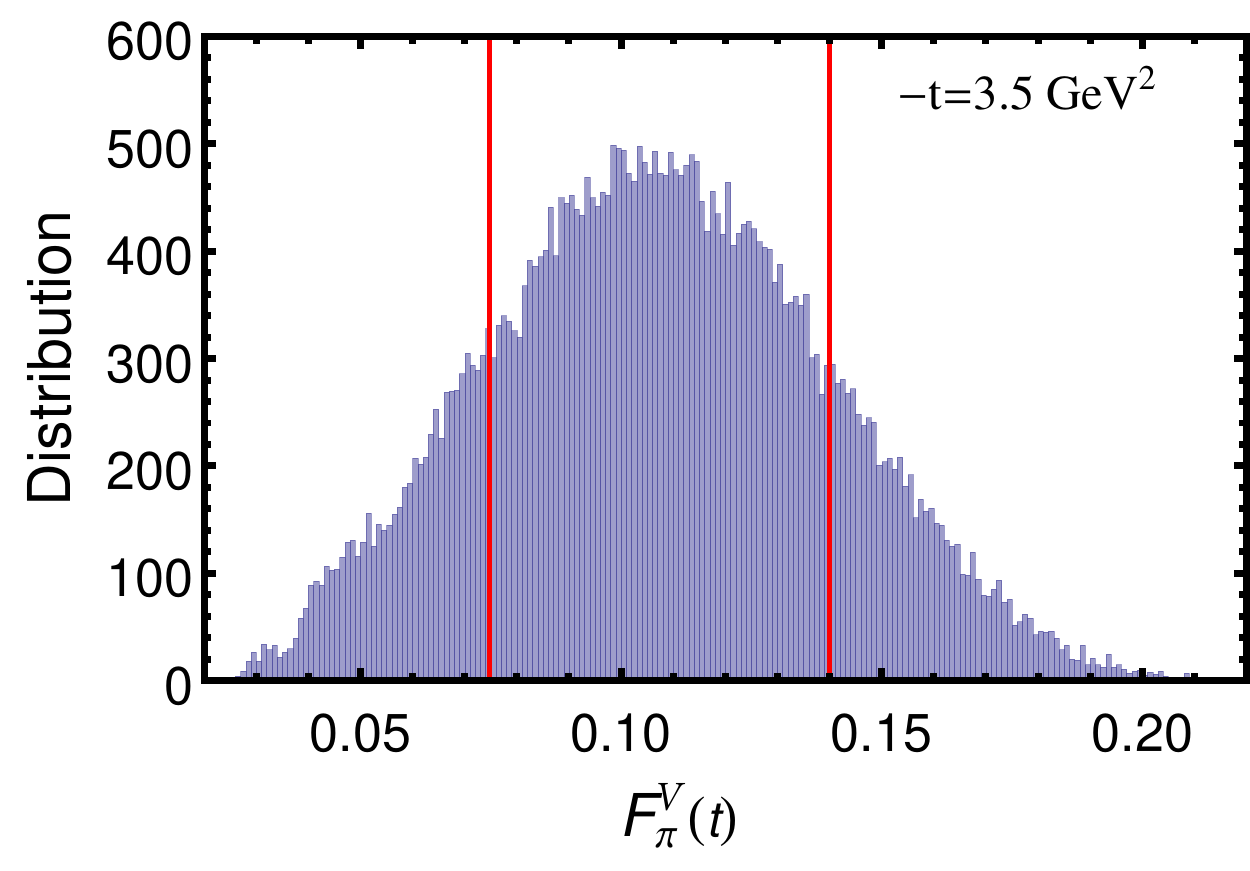}\\
		\includegraphics[width = 8cm]{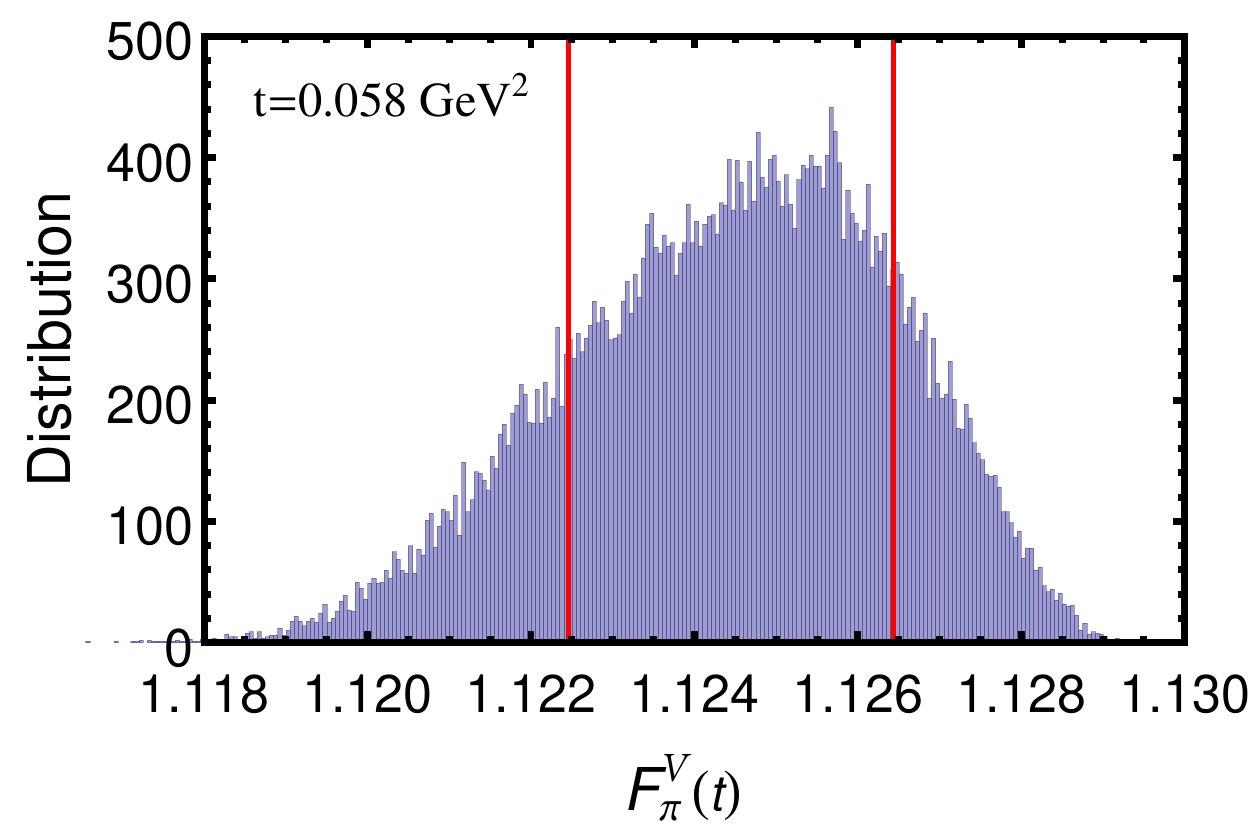}
		\includegraphics[width = 8cm]{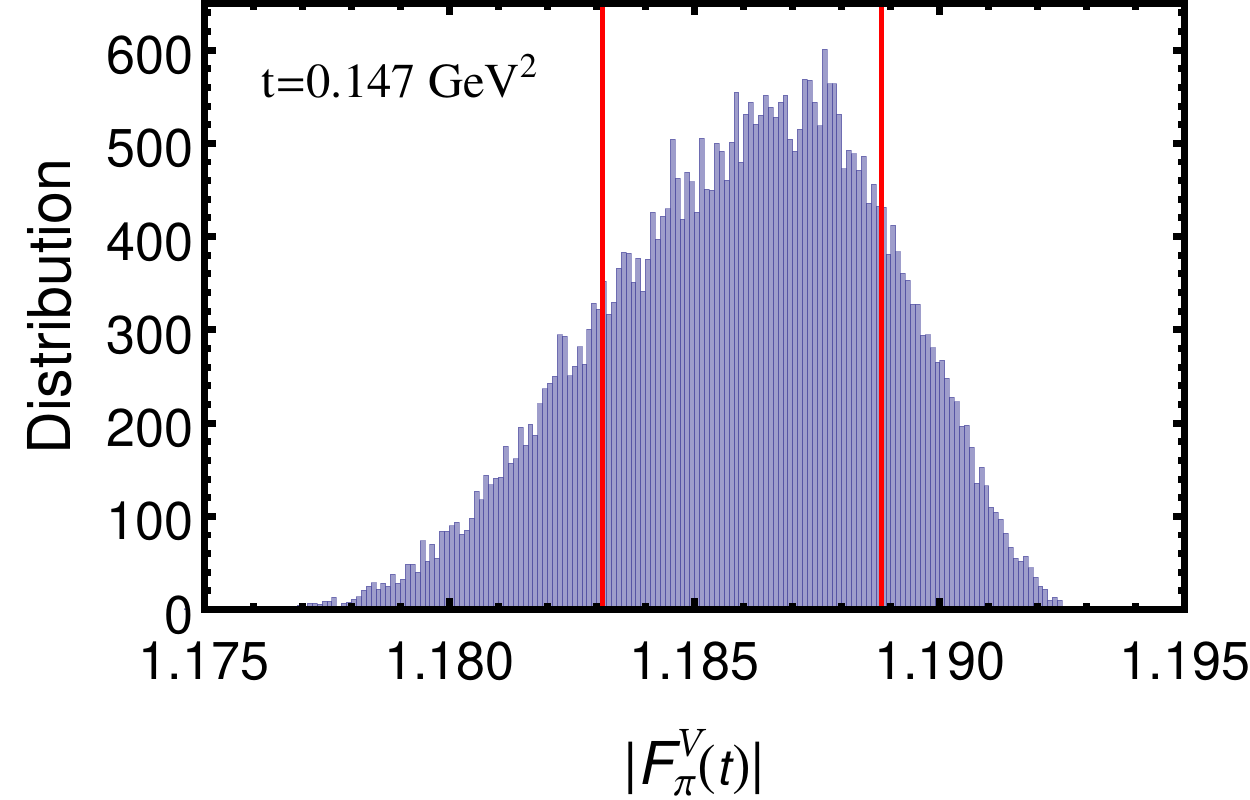}
		\caption{Statistical distributions of the output values of the form factor at two spacelike points (upper panel) and two timelike points, one below and the other above the unitarity threshold $t_+$ (lower panel). In the calculation, we used  the Bern phase, the input from the spacelike point $t_s=-1.6 \gev^2$, and the modulus at the timelike point  $\sqrt{t_t}= 0.699  \gev$ measured by {\it BABAR} \cite{Aubert:2009ad}. The vertical lines indicate the 68.3\% confidence limit (CL) intervals).}
	\end{center}
	\vspace{0.3cm}
\end{figure*}

In Fig. \ref{fig:hist}, we present for illustration the distributions of the output values of the form factor at several points of interest (two spacelike points in the upper panel, and two timelike points, one below and the other above the unitarity threshold, in the lower panel). The histograms have been obtained using as input   the Bern phase, the value at the spacelike point $t_s=-1.6 \gev^2$, and the modulus at one timelike point measured by {\it BABAR} \cite{Aubert:2009ad}.
Similar results have been obtained using as input the Madrid phase and other experimental data.  One can see that the distributions are very close to a Gaussian and allow the extraction of the mean value and the standard deviation (defined as the 68.3\% confidence limit (CL) intervals) for the values of interest  $F_\pi^V(t)$ or its modulus.

The next step is to take the  average of the results obtained with input from various measurements.  Since the degrees of correlations between the measurements at different energies are expected to vary from one experiment to another,  we perform first the average of the values obtained with input from each experiment. As argued in \cite{Schmelling:1994pz}, the most robust average of a set of $n$ measurements $a_i$ is the weighted average
\beq\label{eq:av}
\bar{a}=\sum_{i=1}^{n} w_i a_i, \quad w_i=\frac{1/\delta a_i^2}{\sum_{j=1}^{n}1/\delta a_j^2},
\eeq
where $\delta a_i$ is the error of $a_i$.

For the best estimation of the error in the case of unknown correlations,  the prescription proposed in \cite{Schmelling:1994pz} is to define a function $\chi^2(f)$  
\beq\label{eq:chisq}
\chi^2(f) = \sum_{i,j=1}^{n} (a_i- \bar{a})(C(f)^{-1})_{ij}(a_j-\bar{a}) 
\eeq
in terms of the covariance matrix $C(f)$ with elements
\begin{equation}\label{eq:cov}
  C_{ij}=\begin{cases}
    \delta a_i \delta a_i & \quad \quad \text{if $i=j$},\\
    f  \delta a_i \delta a_j &\quad \quad \text{if $i\neq j$}.
  \end{cases}\
\end{equation}
The parameter  $f$ denotes the fraction of the maximum possible correlation: for $f=0$ the measurements are treated as uncorrelated, for  $f=1$ as fully (100\%) correlated.

If $\chi^2(0)<n-1$, the data might indicate the existence of a positive correlation. The prescription proposed in \cite{Schmelling:1994pz} is to increase $f$ until $\chi^2(f)=n-1$.
With the solution $f$ of this equation, the standard deviation $\sigma (\bar a)$ of $\bar a$ is determined from the variance \cite{Schmelling:1994pz}
\beq\label{eq:sigma}
\sigma^2 (\bar a)=\left (\sum_{i,j=1}^n (C(f)^{-1})_{ij}\right)^{\!\!-1}.
\eeq 
On the other hand, a value $\chi^2(0)>n-1$ is an  indication that the individual errors are underestimated.  If the ratio $\chi^2(0)/(n-1)$ is not very far from 1, the procedure suggested in \cite{Schmelling:1994pz, PDG} is to rescale the variance $\sigma^2 (\bar a)$ calculated with (\ref{eq:sigma})  by the factor $\chi^2(0)/(n-1)$.

 In our work, this procedure was applied first for combining the results obtained with a definite input phase, a specified input (\ref{eq:Huber}) from the spacelike region,  and different measurements in the timelike region available from each experiment listed in Table \ref{tab:1}. In most cases, a large degree of error correlation between the results obtained with different timelike energies was found, as indicated by values close to 1 of the parameter $f$ derived from data. Then the results obtained with the two phases, Bern and Madrid,   were combined in a conservative way by taking the simple average of the central values and of the uncertainties. The same conservative average was used for combining the results obtained with the two spacelike data  (\ref{eq:Huber}).

 The last step was to combine  the individual values obtained with  measurements by the different  experiments listed in Table \ref{tab:1}. Again, the error correlation for these values is difficult to assess {\it a priori}.  Therefore, we have applied the same data-driven procedure described above  for finding  the correlations. Since, as discussed in \cite{Ananthanarayan:2016mns}, the data from $e^+e^-$-annihilation and $\tau$-decay experiments are consistent in the region $0.65-0.71$ GeV, the results from all the 10 experiments in Table \ref{tab:1} can be combined into a single central value and  standard deviation which we quote as the error. 

\section{Results \label{sec:results}}
We have applied the procedure described above for deriving central values and standard deviations for $F_\pi^V(t)$ in three energy regions: small spacelike momenta $Q^2=-t\le 0.25 \gev^2$, where measurements are available from NA7 experiment \cite{Amendolia:1986wj}, larger spacelike momenta, up to $Q^2\leq 8.5 \gev^2$, and the unphysical timelike region $0<t< t_+$ below the unitarity threshold. We have also derived central values and standard deviations for the modulus $|F_\pi^V(t)|$ in the region above the unitarity threshold, below $\sqrt{t}=0.63 \gev$, and have used these results for a new determination of the HVP contribution from energies below $0.63 \gev$ to the   muon $g-2$. In the following subsections we present the results for each kinematical region, namely small spacelike momenta,
large spacelike momenta, unphysical timelike momenta, 
and timelike momenta on the unitarity cut below 0.63 GeV.
The implications of these determinations are also studied in
each of these subsections.

\subsection{Small spacelike  momenta}\label{sec:smallsl}

At small spacelike momenta squared, $Q^2\le 0.25 \gev^2$, the pion form factor has been measured  from $ep$ elastic scattering by the NA7 experiment \cite{Amendolia:1986wj}, considered for a long time  a landmark experiment. Recently, the ETM collaboration \cite{ETM} reported the most precise lattice calculations of $F_\pi^V(-Q^2)$ for small $Q^2$. The comparison with the lattice results has been actually the main motivation for choosing this kinematical region in our study. It turns out that  our predictions for the form factor in this region are very precise: the errors, obtained by the procedure described in the previous section, vary from 0.0005 near the origin to 0.003 at the end  of the region. 

\begin{figure}[thb]\vspace{0.3cm}
		\includegraphics[width = 8cm]{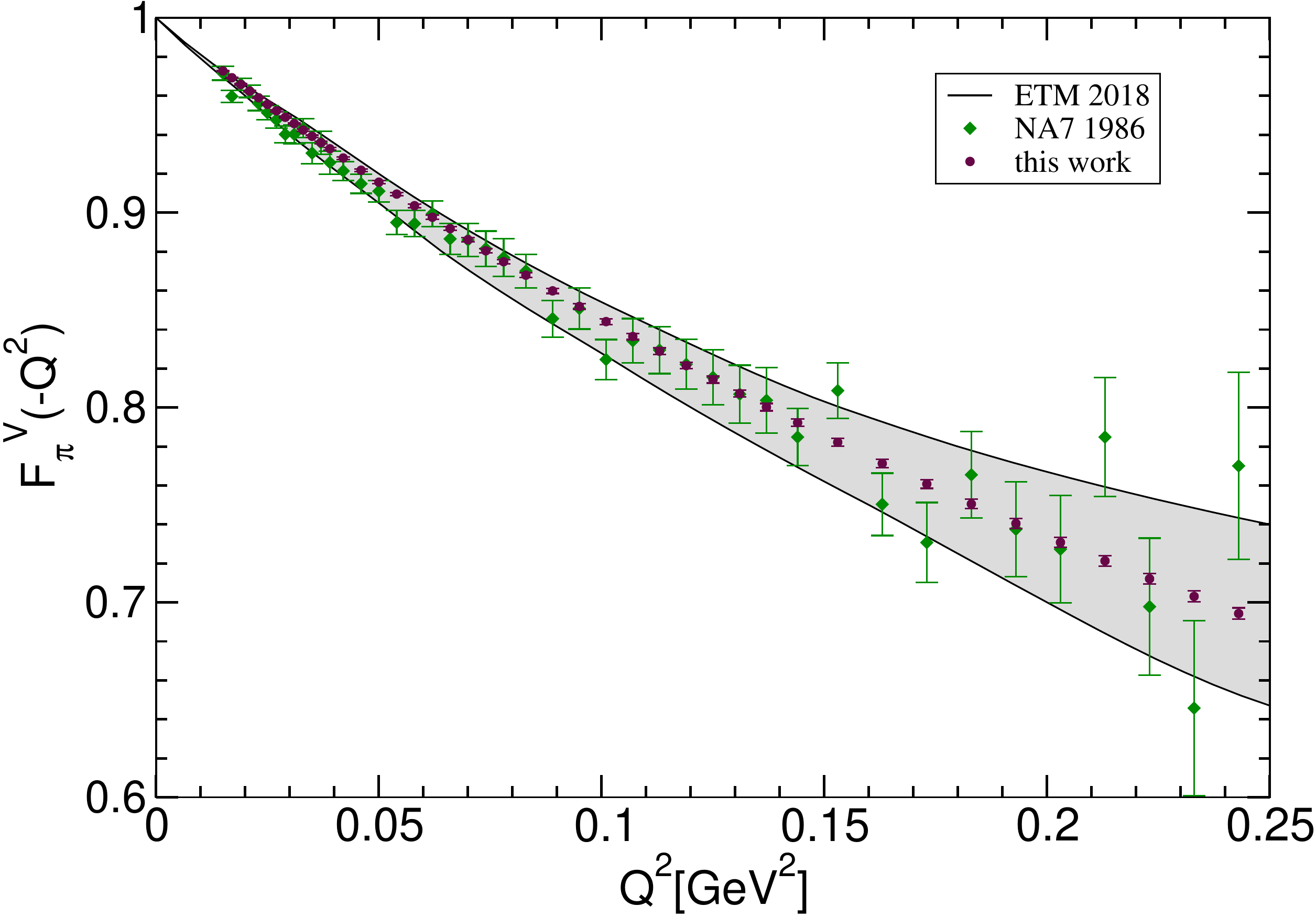}
		\caption{The predictions for the pion form factor in the spacelike region near the origin derived in this work, compared with the experimental results of the NA7 Collaboration \cite{Amendolia:1986wj}  and the lattice calculations of the ETM Collaboration \cite{ETM}.\label{fig:NA7}}
	\vspace{0.3cm}
\end{figure}

 In Fig. \ref{fig:NA7}, we present the values of the form factor calculated in this work at a number of spacelike points below $0.25 \gev^2$.  Also shown are the experimental data from  Ref. \cite{Amendolia:1986wj} and the results of the lattice calculation reported in  Ref.  \cite{ETM}, shown as a band which includes all the uncertainties. One can see that our results are consistent with the lattice values, and are much more precise. It is a challenge for the future lattice calculations to increase the precision to the level reached by the phenomenological determination based on analyticity and unitarity.

It may be noted that our procedure can be extended further as there is no real constraint
on the range of values to be probed in this sector, but for practical purposes, our determination
has been limited to the same range as in the lattice study and in the NA7 experiment.

\subsection{Large spacelike momenta and the onset of perturbative QCD}\label{sec:largesl}

\begin{figure*}[thb]\vspace{0.2cm}
		\includegraphics[width = 8cm]{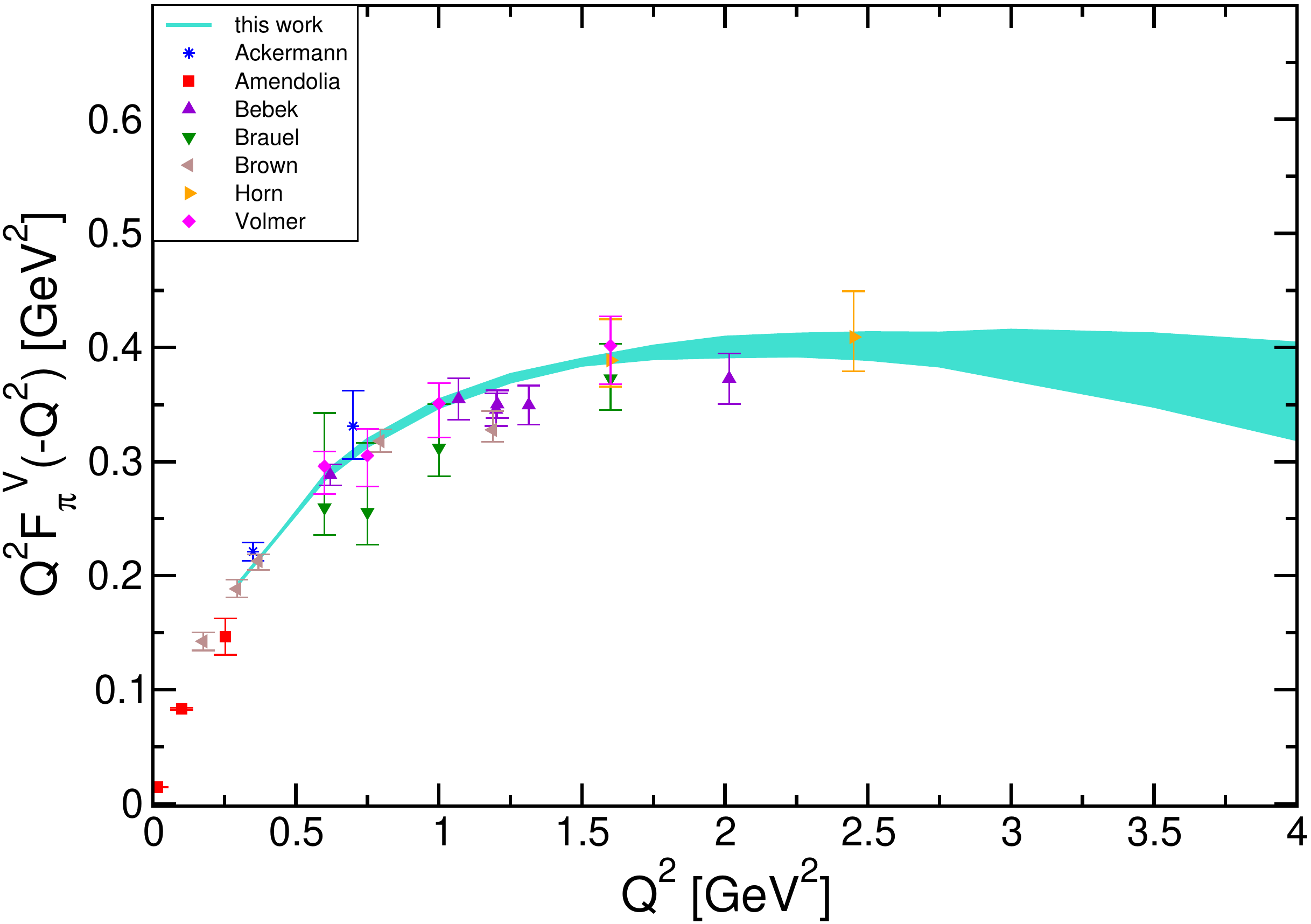}
		\includegraphics[width = 8cm]{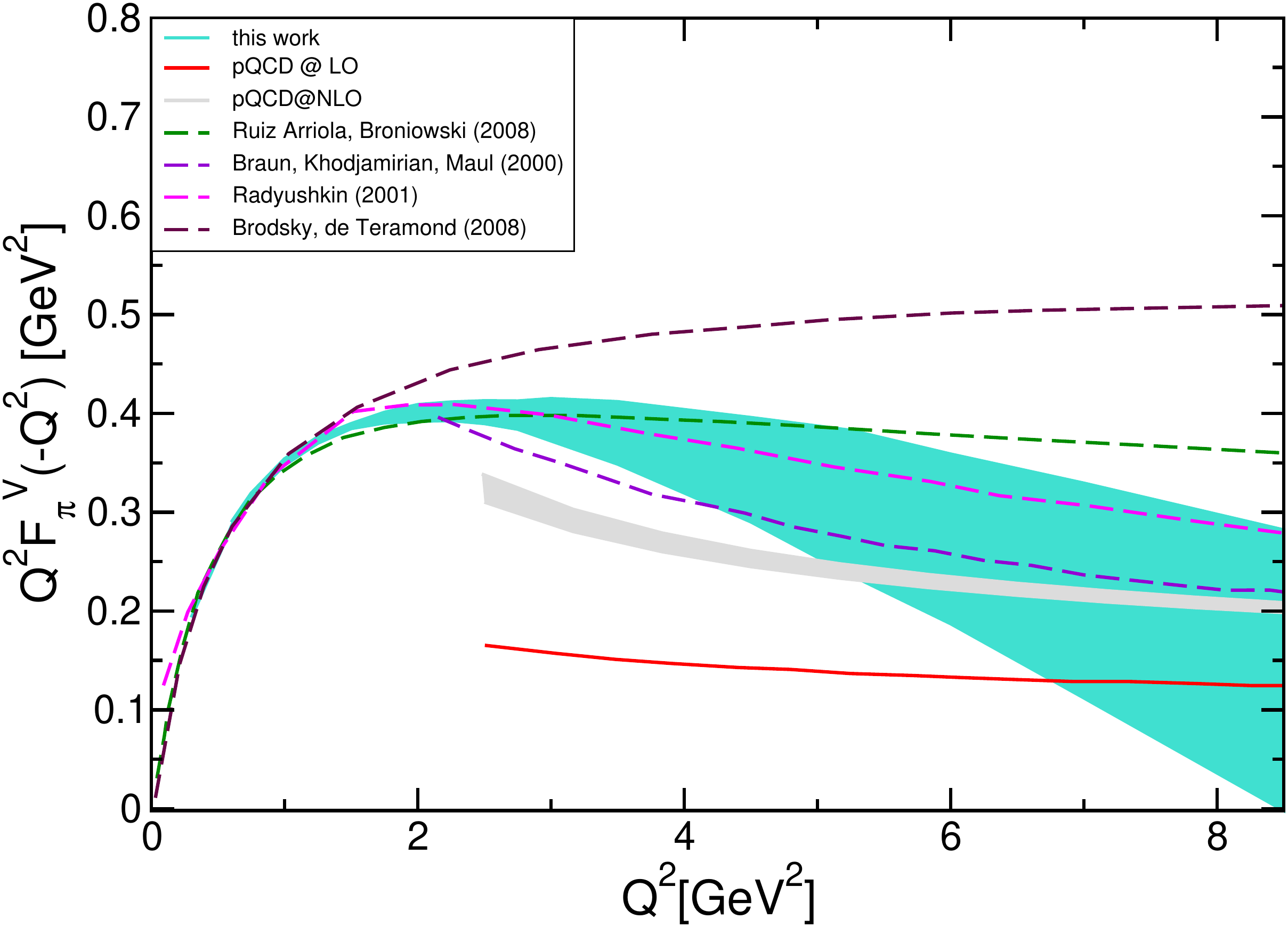}
		\caption{The pion form factor calculated in this work on the spacelike axis, represented as a band which includes the total error. Left panel: comparison with experimental data. Right panel: comparison with perturbative QCD at LO and NLO, and with several nonperturbative models. \label{fig:sl}}
	\vspace{0.3cm}
\end{figure*}

It has long been known that in the case of the pion form factor the asymptotic regime  described by the dominant term  (\ref{eq:qcd}) of perturbative QCD
sets in quite slowly, due to the complexity of soft, nonperturbative processes in QCD in the intermediate
$Q^2$ region. Several nonperturbative approaches have been proposed for the study of the pion form factor, 
including QCD sum rules \cite{Ioffe:1982}, quark-hadron local duality  
\cite{Nesterenko:1982,Radyushkin:1995, Radyushkin:2001,Braguta:2007fj}, 
extended vector meson dominance \cite{Dominguez},
light-cone sum rules 
\cite{Braun:1994,Braun:1999uj, Bijnens:2002}, sum rules with nonlocal condensates 
\cite{Bakulev:1991ps,Bakulev:2004cu,Bakulev:2009ib}, AdS/QCD models \cite{Grigoryan:2007,Brodsky:2007hb}, $k_T$ factorization method \cite{Cheng:2015qra}, dispersion relations treatment \cite{Gorchtein:2011xe}, covariant spectator theory \cite{Biernat:2013aka}, and Dyson-Schwinger equation framework \cite{Chang:2013nia}.  
In particular, the  onset of the asymptotic regime in the presence of Sudakov corrections \cite{Gousset:1994yh}
and large $N_c$ Regge approaches \cite{RuizArriola:2008sq} is expected to be quite slow.
Constructing a fully valid model to describe the form factor at intermediate
energies in fundamental QCD still remains a major theoretical challenge. 

 Measurements of the spacelike form factor for spacelike momenta are reported in  Refs.  \cite{Brown:1973wr,  Bebek:1974iz, Bebek:1976ww, Ackermann:1977rp, Bebek:1978, Brauel:1979zk, Volmer:2000ek, Tadevosyan, Horn, Huber}, the most precise being the recent results of 
the JLab collaboration \cite{Horn, Huber} quoted in Eq. (\ref{eq:Huber}). The lack of precise experimental data in the higher $Q^2$ region 
is a major obstacle to confirm or discard the theoretical models available. 
The calculation presented in this work provides an alternative way for testing the onset of the asymptotic QCD regime and the validity of various theoretical models proposed for intermediate energies.

In the left panel of Fig. \ref{fig:sl}, the predictions of this work for $Q^2<4 \gev^2$, represented as a cyan band which includes the full error, are compared with some of the experimental data. We recall that in our calculation the only input from the spacelike axis consists of the points  given in Eq. (\ref{eq:Huber}), denoted as  Horn   in Fig.  \ref{fig:sl}.  The increased precision of our determinations is due to the timelike information. One can see that, except for a few  points, the experimental measurements are in general agreement with our determinations.

At higher spacelike momenta, the precision of our predictions starts to diminish, since the extrapolation is more sensitive to the values of the form factor at intermediate timelike energies, for which no precise information is  available. To account for this, we
have adopted the conservative, weaker condition (\ref{eq:L2}). Up to $Q^2$ around  $8 \gev^2$, the precision  nevertheless is
{acceptable, allowing us to probe the onset of the asymptotic regime predicted by factorization and perturbative QCD.
In the right panel of  Fig. \ref{fig:sl},  we  compare our predictions shown in cyan band with perturbative QCD at LO and NLO, and with some theoretical models. The gray band corresponds to the NLO result obtained by varying the renormalization scale in suitable range following \cite{Melic:1999mx}

At first sight we note that perturbative QCD at LO can not reliably describe the form factor at $Q^2 \leq 7$~GeV$^2$. Though the description improves at NLO it is still unreliable for $Q^2 \leq 5.5$~GeV$^2$. We limit ourselves only to these conservative statements, since precisely at the energies where the NLO and our predictions start to become compatible, our procedure meets its natural limitations.  This can be seen in the fact that our band hits the $x$-axis in right panel of Fig. \ref{fig:sl},  while there are strong arguments (cf. for instance Ref. \cite{Leutwyler:2002hm}) that this form factor cannot have zeros on the spacelike axis.  Therefore, we refrain from making definite statements for higher $Q^2$, in view of the fact that this is the region where our method lacks the precision that it has in the other three regimes considered in this work.

As we discussed above, there are many theoretical models in the literature for
addressing the properties of the form factors in this region.
For illustration, we have considered the predictions
from four of these as typical examples. For instance, the theoretical models proposed in \cite{Radyushkin:2001,Braun:1999uj} appear to be consistent with the phenomenological band, while the predictions of \cite{Brodsky:2007hb, RuizArriola:2008sq} appear to be too high.

We note finally that the results derived in this work are consistent with those derived in our previous work \cite{Ananthanarayan:2012tn}, being more precise, since we now included information on the modulus of the form factor on the timelike axis and used extensive Monte Carlo simulations for the error analysis.  

\subsection{Unphysical timelike region}\label{eq:unphys}
No experimental information or QCD lattice calculations are available for the pion form factor in the unphysical timelike region between the origin and the unitarity threshold $t_+$. 
For this region our method allows to make very precise predictions. In Table \ref{table:TLbelowcut}, we list the central values and the errors on  $F_\pi^V(t)$
 at several unphysical timelike points.  This region cannot be accessed by
 experiment, but it can be  by the lattice, in principle, so our results can be viewed as a benchmark for lattice predictions in the  future. 

	\begin{table}[hb]\vspace{0.3cm}
		\begin{tabular}{c c  }\hline \hline
			$\sqrt{t}$ GeV	& ~~$F_\pi^V(t)$ \\\hline
			0.140  & ~ $ 1.037 \pm 0.001 $ \\  0.197  & ~ $ 1.078 \pm 0.001 $  \\ $ 0.242 $ & ~ $ 1.124 \pm 0.001 $
			 \\ $ 0.279 $ & ~ $ 1.176 \pm 0.002 $ \\
			\hline\hline
		\end{tabular}\caption{Central values and errors on $F_\pi^V(t)$ in the timelike region below the unitarity threshold $\sqrt{t_+}=2m_\pi$. \label{table:TLbelowcut}}
	\end{table}
In this region the predictions of chiral perturbation theory are expected to be most accurate. The precise determinations in table \ref{table:TLbelowcut} can be used to determine the curvature $c$ and higher shape parameter $d$ of the Taylor expansion $F^V_\pi(t)= 1+ \langle r_\pi^2 \rangle t/6 + ct^2 + dt^3 +\mathcal{O}(t^4)$. 

\subsection{Low energies above the unitarity threshold and the contribution to muon $g-2$}\label{sec:tl}
As mentioned in the Introduction, above the unitarity threshold, where the form factor is a complex function, its modulus is extracted from  the cross section of the $e^+e^-\to \pi^+\pi^-$ process, or, using isospin symmetry, from the hadronic decay of the $\tau$ lepton.  The $\tau$ decay  has been for a long time the most precise source of information, in spite of the nontrivial corrections that are required to convert the measured spectral functions into genuine values of $|F_\pi^V(t)|$. However, the accuracy of the   $e^+e^-$ experiments improved gradually, the extraction of the modulus being based at present almost exclusively on them. 

At low energies, the modulus of the form factor is poorly known, due to the difficulties of the experimental measurements in this region: only two experiments, {\it BABAR} \cite{Aubert:2009ad} and KLOE \cite{Ambrosino:2008aa, Ambrosino:2010bv, Babusci:2012rp} 
reported data at low energies, and unfortunately they are not consistent among them. Our method allows a precise determination of $|F_\pi^V(t)|$ at low energies.  In Fig. \ref{fig:TLcut} we present our results, together with the experimental values of {\it BABAR} \cite{Aubert:2009ad} and KLOE \cite{Ambrosino:2010bv, Babusci:2012rp}.
For convenience, we show the  values  of the modulus squared, which enter directly into the calculation of the two-pion contribution to the muon magnetic anomaly. 
One can see that our predictions are much more precise than the available experimental results, especially at energies below 0.5 GeV. 

\begin{figure}[thb]\vspace{0.3cm}
		\includegraphics[width = 8cm]{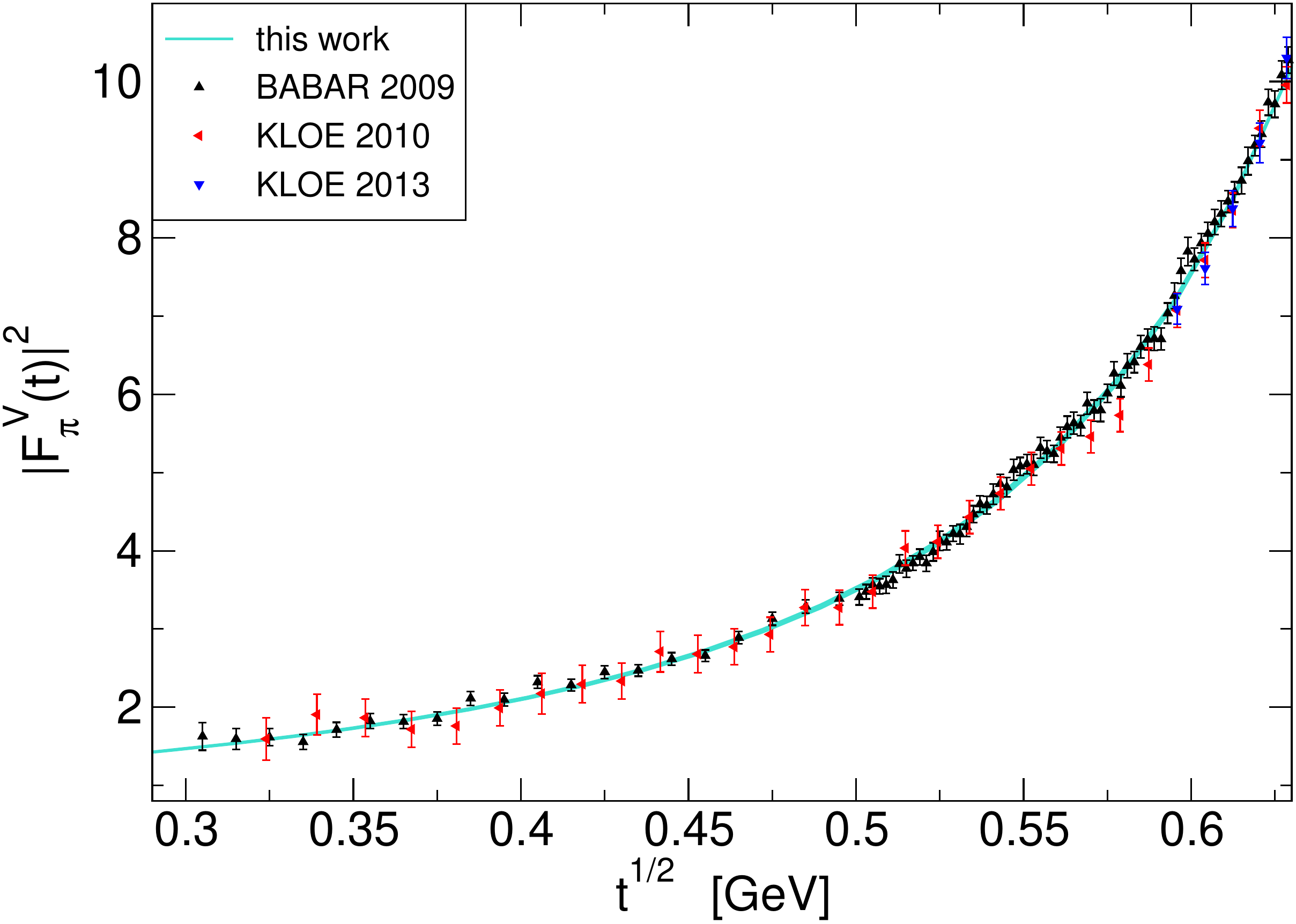}
		\caption{Our predictions for the modulus squared of the pion form factor on the cut below 0.63 GeV, compared with {\it BABAR} and KLOE experimental data.\label{fig:TLcut}}
	\vspace{0.3cm}
\end{figure}

For completeness, we list in Table \ref{tab:TLcut} the  central values and the uncertainties of the modulus squared of the form factor at several energies below $0.63 \gev$.

\begin{table}[thb]\vspace{0.3cm}
		\begin{tabular}{c c c c  }\hline \hline
			$\sqrt{t}$ GeV	& ~~$|F_\pi^V(t)|^2$ & ~~~ $\sqrt{t}$ GeV & ~~~ $|F_\pi^V(t)|^2$\\\hline
			0.281  & ~ $ 1.389 \pm 0.004 $ & ~~ $ 0.437 $ & ~~ $ 2.485 \pm 0.014 $  \\
			0.283  & ~ $ 1.397 \pm 0.004 $ & ~~ $ 0.455 $ & ~~ $ 2.712 \pm 0.016 $ \\
			0.285  & ~ $ 1.405 \pm 0.004 $ & ~~ $ 0.472 $ & ~~ $ 2.978 \pm 0.019 $ \\
			0.291  & ~ $ 1.431 \pm 0.004 $ & ~~ $ 0.490 $ & ~~ $ 3.291 \pm 0.022 $ \\
			0.297  & ~ $ 1.456 \pm 0.004 $ & ~~ $ 0.507 $ & ~~ $ 3.664 \pm 0.025 $ \\
			0.314  & ~ $ 1.536 \pm 0.005 $ & ~~ $ 0.525 $ & ~~ $ 4.111 \pm 0.028 $ \\
			0.332  & ~ $ 1.626 \pm 0.006 $ & ~~ $ 0.542 $ & ~~ $ 4.653 \pm 0.031 $ \\
			0.349  & ~ $ 1.728 \pm 0.007 $ & ~~ $ 0.560 $ & ~~ $ 5.318 \pm 0.034 $ \\
			0.367  & ~ $ 1.842 \pm 0.008 $ & ~~ $ 0.577 $ & ~~ $ 6.144 \pm 0.036 $ \\
			0.384  & ~ $ 1.972 \pm 0.009 $ & ~~ $ 0.595 $ & ~~ $ 7.174 \pm 0.031 $ \\
			0.402  & ~ $ 2.120 \pm 0.011 $ & ~~ $ 0.612 $ & ~~ $ 8.498 \pm 0.028 $ \\
			0.419  & ~ $ 2.290 \pm 0.012 $ & ~~ $ 0.630 $ & ~ $ 10.177 \pm 0.005 $ \\
			\hline\hline
		\end{tabular}\caption{Central values and errors on $|F_\pi^V(t)|^2$ in the range from two-pion threshold to 0.63 GeV. \label{tab:TLcut}}
\end{table}

We shall use now these results for making a new determination of the low-energy pion-pion contribution to the muon anomaly.  
The leading order (LO) two-pion contribution to $a_\mu$ from energies below $\sqrt{t_{up}}$, which does not contain the
vacuum polarization  but includes one-photon final-state radiation (FSR),  is expressed in terms of  $F_\pi^V(t)$ as
\begin{equation} \label{eq:amu}
a_\mu^{\pi\pi}|_{\leq \sqrt{t_{up}}} = \frac{\alpha^2 m_\mu^2}{12 \pi^2}\int_{t_+}^{t_\up} \frac{dt}{ t} K(t)\ \beta^3_\pi(t) \  F_{\rm FSR}(t)\
|F_\pi^V(t)|^2.\end{equation}
In this relation,  $\beta_\pi^3(t)=(1-4 m_\pi/t)^{3/2}$ is the two-pion phase space relevant for $e^+ e^-\to \pi^+\pi^-$ annihilation, 
\begin{equation} \label{eq:K}
K(t) = \int_0^1 du(1-u)u^2(t-u+m_\mu^2u^2 )^{-1}
\end{equation}
is the QED kernel function  \cite{CzMa}, which exhibits a drastic increase at low $t$, and
\beq
 F_{\rm FSR}(t)=\left(1+\frac{\alpha}{\pi}\,\eta_\pi(t)\right)
\eeq
 is the FSR correction, calculated in scalar QED \cite{Czyz:2004rj, Bystritskiy:2005ib}.

Using the central values of $|F_\pi^V(t)|^2$ given in Table \ref{tab:TLcut}, the integral (\ref{eq:amu}) gives $(132.97 \pm 0.07)  \times 10^{-10}$, where we quoted an uncertainty due to the method of integration. In order to estimate the statistical error $\sigma_{a_\mu}$ of this result, we shall apply the  standard error propagation, expressed in our case as
\beq\label{eq:error}
\sigma_{a_\mu}=\left[\int_{t_+}^{t_\up}\int_{t_+}^{t_\up} dt dt'\ {\rm Cov}(t,t')\ W(t) W(t')\right]^{1/2},
\eeq
where
\beq\label{eq:W}
W(t) =\frac{\alpha^2 m_\mu^2}{12 \pi^2} \ \frac{K(t) }{ t} \ \beta^3_\pi(t) \  F_{\rm FSR}(t),
\eeq
and ${\rm Cov}(t,t')$ is the covariance matrix describing the correlation of the errors on $|F_\pi^V|^2$ at two points $t$ and $t'$. For a most conservative estimate, we  assume fully correlated errors, which means that we take
\beq
{\rm Cov}(t,t')=\sigma(t) \sigma(t'), 
\eeq
where $\sigma(t)$ is the error on $|F_\pi^V(t)|^2$, determined by the procedure described in Sec. \ref{sec:ff}. Then  the integral (\ref{eq:error}) gives an error of $0.69 \times 10^{-10}$. Adding to this the integration error quoted above, we finally obtain
\beq\label{eq:res}
a_\mu^{\pi\pi}|_{\leq 0.63\gev} = (132.97\pm 0.70)\times 10^{-10}.
\eeq

For further comparison, we quote also the result obtained using the timelike input on the modulus in the range (0.65-0.71) GeV only from the $e^+e^-$ experiments:
\beq\label{eq:res1}
a_\mu^{\pi\pi}|_{\leq 0.63 \gev} = (132.91\pm 0.76)\times 10^{-10}.
\eeq

The values (\ref{eq:res}) and (\ref{eq:res1}) are fully consistent with our previous results $(133.26 \pm 0.72 )\times 10^{-10}$ and  $(133.02 \pm 0.77)\times 10^{-10}$, respectively, obtained in \cite{Ananthanarayan:2016mns} for the same quantities with a slightly different method. The difference stems from the fact that in Ref.~\cite{Ananthanarayan:2016mns} we generated the statistical distribution of the integral (\ref{eq:amu})  directly from Monte Carlo simulations,  without deriving the modulus squared of the form factor  at each energy below 0.63 GeV.

We quote also the result $ a_\mu^{\pi\pi}|_{\leq 0.63 \gev} =132.5(1.1)\times 10^{-10}$  of the recent analysis \cite{Colangelo:2018}, which exploits  analyticity and unitarity by using an extended Omn\`es representation of the form factor in a global fit of the phenomenological data on $e^+e^-\to \pi^+\pi^-$ from energies below 1 GeV and the NA7 experiment.
We note also that the direct integration of the interpolation of the  $e^+e^-$ data below 0.63 GeV, proposed in \cite{Teubner:2018}, gives\footnote{We thank T. Teubner for  this calculation.} $a_\mu^{\pi\pi}|_{\leq 0.63 \gev} = (131.12 \pm 1.03)\times 10^{-10}$.

It may be noted that the explicit values listed in Table \ref{tab:TLcut} for this region allow an evaluation of the dipion contribution to the muon anomaly in any interval of interest.

\section{Discussion and Conclusions\label{sec:conclusion}}


 In the present work we have obtained high-precision 
 predictions for the pion electromagnetic form factor in several kinematical regions of interest.  We have used a
method based on analyticity and unitarity, which does not involve standard dispersion relations or specific parametrizations. The input, summarized in Sec. \ref{sec:aim}, consists of the phase of the form factor on a part of the unitarity cut and a conservative integral condition on the modulus squared on the remaining part. The experimental values at some discrete points on the timelike and the spacelike axes are also included. 

Using the solution of a functional extremal problem formulated in Sec. \ref{sec:aim} and discussed in the Appendix, we have derived optimal upper and lower bounds on the values of the form factor (or its modulus) in the regions of interest, which are expressed only in terms of the adopted input and involve no model-dependent assumptions. 
A key element of our method is the determination of the central values and the errors from statistical distributions obtained from a large set of pseudo-data, and the conservative combination of the results from various experiments using data-driven error correlations.  We emphasize that, since we do not use a specific parametrization for the form factor and the analytic bounds do not involve model-dependent assumptions, there are no additional systematic errors in our approach. Thus,  the Monte Carlo simulations described in Sec. \ref{sec:ff} give the full errors of our predictions for the form factor. 

  We mention that the same technique has been applied in  \cite{Ananthanarayan:2017efc} for a precise extraction of the pion electromagnetic charge radius, and in
 \cite{Ananthanarayan:2016mns} for a direct calculation of the two-pion low-energy contribution to muon $g-2$.
 
The high-precision determinations of the form factor (or its modulus) in several significant  kinematical regions are presented in Sec. \ref{sec:results}. In particular, on the spacelike axis at low $Q^2$ our results are much more precise than the recent lattice calculations \cite{ETM}, and at larger  $Q^2$ we confirm our previous conclusion \cite{Ananthanarayan:2012tn} that the asymptotic regime of perturbative QCD is away from the region $Q^2\le 7 \gev^2$.

On the timelike axis, we derived  high-precision values of the modulus squared of the form factor on the unitarity cut below $0.63 \gev$, shown in Fig. \ref{fig:TLcut} and Table \ref{tab:TLcut}. Our predictions are much more precise than the experimental values available in this region from {\it BABAR} and KLOE experiments, especially below 0.5 GeV. Also, in Sec.~\ref{sec:results}, the determinations we provide in the unphysical timelike region could serve as a benchmark for theoretical probes in this region.

From the precise values given in Table \ref{tab:TLcut}, we have performed  a new determination of the two-pion contribution from low energies to the muon $g-2$. Our results for $a_\mu^{\pi\pi}|_{\leq 0.63 \gev}$ are given in Eqs. (\ref{eq:res}) and  (\ref{eq:res1}), where the first uses the input from both $e^+e^-$ and $\tau$ experiments, and the second only from  $e^+e^-$  experiments.
These results are consistent with the values derived in our previous work \cite{Ananthanarayan:2016mns},  where the technique of rigorous analytic bounds and Monte Carlo simulations has been applied in a slightly different way, by deriving a statistical distribution directly for the quantity  $a_\mu^{\pi\pi}|_{\leq 0.63 \gev}$. 

As seen from the values quoted at the end of the previous section, our results are consistent with the prediction of the recent analysis \cite{Colangelo:2018} based on analyticity and unitarity, while the result obtained from the direct integration of the data \cite{Teubner:2018} is slightly lower.  We emphasize that we do not use as input experimental data from energies below 0.63 GeV or from NA7 experiment.
Thus, our prediction for $a_\mu^{\pi\pi}|_{\leq 0.63 \gev}$ is to a certain extent complementary to the determination  of the analysis performed in \cite{Colangelo:2018},  which exploits analyticity and unitarity in a different way and uses as input low-energy data. 

This work represents
the state of the art in an important low-energy sector of the Standard Model, which is
going to be tested at the upcoming Fermilab experiment E969. In contrast to  our prior publications \cite{Ananthanarayan:2017efc,Ananthanarayan:2016mns}, which were focused on the determination of a single number, here we have obtained an extensive tabulation of the values of the electromagnetic
form factor in several significant kinematical regions.
Using these determinations, the value of the dipion contribution to the muon anomaly
remains consistent with the value reported earlier, proving the robustness of  the approach.

The present work combines strong theoretical inputs with modern Monte Carlo methods
along with high precision experimental information and phase
shift information in regions where experiments are in agreement
to shed light on regions where either experiments do not have
sufficient precision or where there are significant disagreements,
or regions which are not directly accessible by experiment.
It also offers a test to theoretical predictions based on very
different approaches to strong interaction phenomenology.

\subsection*{Acknowledgments}We would like to thank Urs Wenger for useful correspondence. B.A. acknowledges support from the MSIL
Chair of the Division of Physical and Mathematical Sciences, Indian Institute of Science. I.C. acknowledges support from the   Ministry of Research and Innovation of Romania, Contract No. PN 18090101/2018. D.D. is supported by the DST, Government of
India, under INSPIRE Fellowship (No. DST/INSPIRE/04/2016/002620).


\appendix

\section{Solution of the extremal problem }\label{sec:A}

Using the approach proposed in \cite{Caprini:1999ws}, the extremal problem formulated at the end of  Sec. \ref{sec:aim} can be reduced to a standard analytic interpolation problem  \cite{Duren} (also known as a Meiman problem \cite{Meiman}). We review in what follows the main steps of the proof. As discussed in Sec. \ref{sec:input}, we first remove from the form factor the isospin-violating correction $F_\omega(t)$, so in what follows we shall consider  the 
function $F(t)$ defined in (\ref{eq:F}). 
The next step is to introduce a function $h(t)$  by writing
\beq\label{eq:h}
F(t)=\omnes(t) h(t),
\eeq
where $\omnes(t)$ is  the Omn\`es function defined as
\beq	\label{eq:omnes}
 \omnes(t) = \exp \left(\D\frac {t} {\pi} \int^{\infty}_{t_+} dt' 
\D\frac{\delta (t^\prime)} {t^\prime (t^\prime -t)}\right).
\eeq
In this relation, $\delta(t)$ is equal to $\delta_1^1(t)$  at
$t\le \tin$ and is an arbitrary smooth (Lipschitz continuous) function above $\tin$,
which approaches asymptotically $\pi$. 

From the Fermi-Watson theorem (\ref{eq:watson}), it follows that  $h(t)$
is real on the real axis below $\tin$, since the phase of $F(t)$ is exactly compensated by the phase of  $\omnes(t)$. Taking into account the fact that $h(t)$ satisfies the Schwarz reflection property, this implies that it is holomorphic on the real axis below $\tin$, having a branch cut only for $t\ge \tin$. 

In terms of $h(t)$, the equality (\ref{eq:L2}) can be written as
\beq\label{eq:hL2}
\D\frac{1}{\pi} \int_{\tin}^{\infty} dt\, 
\rho(t) |\omnes(t)|^2 |h(t)|^2 \leq  I.
\eeq
 This relation can be written in a canonical form if we perform the conformal transformation,
 \beq\label{eq:ztin}
 \tilde z(t) = \frac{\sqrt{\tin} - \sqrt {\tin -t}} {\sqrt{\tin} + \sqrt {\tin -t}},
 \eeq
and express the factors multiplying  $|h(t)|^2$ in terms of an outer function, 
{\it i.e.} a function analytic and without zeros in 
the unit disk $|z|<1$. In practice, it is convenient to construct it as a product of two outer functions  \cite{Caprini:1999ws, Abbas:2010jc}: the first one, denoted as $w(z)$, has the modulus equal to  
$\sqrt{\rho(t)\, |{\rm d}t/ {\rm d} \tilde z(t)|}$. 
For the choice $\rho(t)=1/t$, it is given by the simple expression 
\beq\label{eq:outerfinal0}
w(z)=\sqrt{\frac{1-z}{1+z}}.
\eeq 
The second outer function, denoted as $\omega(z)$,  
  has the modulus equal to  $|\omnes(t)|$, and can be calculated by the integral representation
\beq\label{eq:omega}
 \omega(z) =  \exp \left(\D\frac {\sqrt {\tin - \tilde t(z)}} {\pi} \int^{\infty}_{\tin}  \D\frac {\ln |\omnes(t^\prime)|\, {\rm d}t^\prime}
 {\sqrt {t^\prime - \tin} (t^\prime -\tilde t(z))} \right).
\eeq 

If we define the function $g(z)$ by
\beq	\label{eq:gF11}
 g(z) = w(z)\, \omega(z) \,h(\tilde t(z)),
\eeq 
where  $\tilde t(z)$ is the inverse of $z = \tilde z(t)$  defined in
Eq.(\ref{eq:ztin}), the condition (\ref{eq:hL2}) can be written with no loss of information as
 \beq\label{eq:gI1}
 \frac{1}{2 \pi} \int^{2\pi}_{0} {\rm d} \theta |g(\zeta)|^2 \leq I, \qquad \zeta=e^{i\theta}.
 \eeq
This condition leads to rigorous correlations 
among the values of the analytic function  $g(z)$ and its derivatives 
at points inside the holomorphy domain, $|z|<1$ (for a proof and earlier references see  Ref. \cite{Abbas:2010jc})  In particular, in our case this amounts to the positivity condition
\beq\label{eq:posit}
{\cal D}\ge 0
\eeq
 of the determinant ${\cal D}$ defined as
\beq\label{eq:det}
{\cal D}=\left|
\begin{array}{c c c c c c}
I-g(0)^2 & \xi_{1} & \xi_{2} &   \xi_{3}\\
	\xi_{1} & \D \frac{z^{2}_{1}}{1-z^{2}_1} & \D
\frac{z_1z_2}{1-z_1z_2}  &  \D  \frac{z_1z_3}{1-z_1z_3} \\
	\xi_{2} & \D \frac{z_1 z_2}{1-z_1 z_2} & 
 \D  \frac{z_2^2}{1-z_2^2}  &  \D  \frac{z_2z_3}{1-z_2z_3} \\
	\xi_3 & \D \frac{z_1 z_3}{1-z_1 z_3} & \D
\frac{z_2 z_3}{1-z_2 z_3} &  \D  \frac{z_3^2}{1-z_3^2} \\
	\end{array}\right|,
\eeq
 where the real values $z_n \in (-1,1)$ are defined as
 \beq\label{eq:zn}
z_n=\tilde z(t_n), 
\eeq
in terms of the two points $t_1=t_s$ and $t_2=t_t$ used as input and the value $t_3$ where we want to calculate bounds on the form factor, and
\beq\label{eq:barxi}
  \xi_n =g(z_n) - g(0), \quad 1\leq n \leq 3.\eeq

The inequality (\ref{eq:posit}) defines an allowed domain for the real values $g(z_n)$.
For $n=1$ and $n=3$ with $t<t_+$,  we have from Eqs. (\ref{eq:h}) and (\ref{eq:gF11})
\beq\label{xin}
g(z_n)=w(z_n)\, \omega(z_n) \,F(t_n) /\omnes(t_n), 
\eeq
while for $n=2$ and $n=3$ with $t>t_+$
\beq\label{eq:gFn} g(z_n) = w(z_n)\, \omega(z_n) \,|F(t_n)| /|\omnes(t_n)|,
\eeq
 where the modulus $|\omnes(t)|$ of the Omn\`es function is obtained from (\ref{eq:omnes}) by the principal value (PV) Cauchy integral
\beq	\label{eq:modomnes}
 |\omnes(t)| = \exp \left(\frac {t} {\pi} \text{\rm PV} \int^{\infty}_{t_+} dt' 
\D\frac{\delta (t^\prime)} {t^\prime (t^\prime -t)}\right).
\eeq

One can show that for each specific input, the determinant (\ref{eq:det}) is a concave quadratic function of the unknown value $F(t_3)$ for $t_3<t_+$, or the modulus $|F(t_3)|$ for $t_3>t_+$, so the inequality (\ref{eq:posit}) can be written as
\beq\label{eq:ABC}
Ax^2+2 B x+C \ge 0, \quad A\leq 0,
\eeq
where $x=F(t_3)$ or $x=|F(t_3)|$.   
This inequality leads to a definite allowed range for $x$ if $B^2-AC\ge 0$ and has no solution if $B^2-AC< 0$. The latter case occurs when the phenomenological input adopted is inconsistent with analyticity. 

From the inequality (\ref{eq:ABC}), one can obtain upper and lower bounds on $F(t_3)$ (or $|F(t_3)|$), expressed in terms of the adopted input. Finally, the isospin correction is applied back according to (\ref{eq:F}), for obtaining the desired bounds on the form factor $F_\pi^V(t)$.

One can prove \cite{Caprini:1999ws, Abbas:2010jc}, that the bounds  are optimal and their values do not depend on the unknown phase of the form factor above the inelastic threshold $\tin$ (the dependence of the Omn\`es function (\ref{eq:omnes}) on the arbitrary phase $\delta(t)$ for $t>\tin$ is compensated exactly by the corresponding dependence of the outer function  (\ref{eq:omega})).  Furthermore, for a fixed weight $\rho(t)$ in (\ref{eq:L2}),  the  bounds  become stronger/weaker when the 
value of the value of $I$ is decreased or increased, respectively. These properties make the formalism model independent and robust against the uncertainties from the high energy region.


\end{document}